\documentclass[american,reprint,aps,pra,superscriptaddress,showpacs]{revtex4-1}
\usepackage{graphics}
\usepackage[T1]{fontenc}
\usepackage[latin9]{inputenc}
\usepackage{babel}
\usepackage{amsthm}
\usepackage{amsmath}
\usepackage{amssymb}
\usepackage[unicode=true,pdfusetitle,
 bookmarks=true,bookmarksnumbered=false,bookmarksopen=false,
 breaklinks=false,pdfborder={0 0 0},backref=false,colorlinks=false]
 {hyperref}
\hypersetup{
 colorlinks,linkcolor=myurlcolor,citecolor=myurlcolor,urlcolor=myurlcolor}
\usepackage{times}
\usepackage{txfonts}
\usepackage{braket}
\usepackage{colortbl}
\definecolor{myurlcolor}{rgb}{0,0,0.7}
\begin{document}
\bibliographystyle{apsrev}
\title{Measuring Quantum Coherence in Multi-Slit Interference}
\author{Tania Paul}
\email{tania2992@gmail.com}
\author{Tabish Qureshi}
\email{tabish@ctp-jamia.res.in}
\affiliation{Centre for Theoretical Physics, Jamia Millia Islamia, New Delhi-110025, India.}


\begin{abstract}
A quantitative measure of quantum coherence was recently introduced,
in the context of quantum information theory. This measure has also been
propounded as a good quantifier of the wave nature of 
quantum objects. However, actually measuring coherence in an experiment
is still considered a challenge. A procedure for measuring 
coherence in a multi-slit interference is proposed here. It can be used for
experimentally testing duality relations for interference experiments
involving more than two slits.
\pacs{03.65.Ta}

\end{abstract}
\maketitle

\section{Introduction}

Wave-particle duality is a very interesting and intriguing aspect of quantum
theory. Bohr argued that the wave and particle aspects of quantum objects,
which we shall call {\em quantons}, are
complementary in nature, in the sense that if an experiment exposes the wave
nature, it will completely hide the particle aspect and vice-versa \cite{bohr}.
Einstein had proposed his famous recoiling slit experiment
(see e.g. \cite{tqeinstein}) in order to refute the complementarity principle.
It was later realized that wave and particle aspects could be revealed 
simultaneously, although to a quantitavely limited extent. Wave-particle
duality is now understood as a constraint on the quantitative measures of
the wave and particle natures, namely duality relations for interference
experiments \cite{wootters,greenberger,englert}. Englert had proved a
duality relation for a two-path interference, which is an inequality
involving a {\em distinguishability}
${\mathcal D}$ and fringe visibility ${\mathcal V}$,
\begin{equation}
{\mathcal D}^2 + {\mathcal V}^2 \le 1.
\label{2duality}
\end{equation}
Distinguishability is a measure which assumes the presence of a device
which is capable of determining which of the two slits, or two paths,
the quanton went through.  If the
device determines which path the quanton has traveled through
without any error (i.e., $\mathcal{D}=1$), then no interference fringes
will appear at the detector (i.e., $\mathcal{V}=0$). On the other hand,
if there is an ambiguity in the which-path information (i.e., $\mathcal{D}\neq
1$), the quanton will show a reduced fringe visibility
(i.e., $\mathcal{V} \neq 0$). The duality relation (\ref{2duality}) was 
also experimentally verified \cite{Durr98}.
Wave-particle duality in two-slit experiments has also been connected to things
like entropic uncertainty relations \cite{coles}, and the dichotomy between
symmetry and asymmetry \cite{vaccaro}.

Wave-particle duality is expected to hold even when a quanton goes through
more than two slits or paths.  Several attempts were made to formulate
a similar duality relation for the case of multi-path experiments
\cite{jaeger,durr,bimonte,englertmb,bimonte1}, without a completely satisfactory
duality relation. The issue of wave-particle duality in multi-path experiments
has also been probed in the framework of entropic uncertainty \cite{colesmb}.
For the particular case of three-slit interference,
a new duality relation was recently derived \cite{3slit}, where the wave nature was 
characterized by the conventional fringe visibility or contrast, but the
particle nature was characterize by a new distinguishability $\mathcal{D}_Q$
which is based on unambigious quantum state discrimination (UQSD)
\cite{uqsd,dieks,peres,jaeger2,bergou}. The duality relation, for three-slit
interference, has the form \cite{3slit}
\begin{equation}
{\mathcal D_Q} + {2{\mathcal V}\over 3- {\mathcal V}} \le 1 .
\label{3duality}
\end{equation}
If one uses this definition of distinguishability, one gets a different form
of two-slit duality relation \cite{3slit}
\begin{equation}
{\mathcal D_Q} + \mathcal{V}\le 1 .
\label{2dualitytq}
\end{equation}
Although different in form, the above relation is completely equivalent
to (\ref{2duality}).

The form of the above two duality relations is different. 
This means that a universal duality relation for
multi-slit interference,
{\em involving fringe visibility}, is probably not possible. A new measure of 
wave nature could probably make a universal duality relation possible.
Coherence in optics has long been thought to be representative of wave
properties, and it has also been connected to distinguishability of paths
\cite{mandelcoherence}. However, a good quantitative measure of quantum
coherence was missing.

Recently a measure of coherence was introduced,
which is just the sum of the absolute values of the off-diagonal elements
of the density matrix of a system namely
$\sum_{i\neq j} |\rho_{ij}|$, 
with $\rho_{i,j}=\langle i | \rho | j \rangle$ \cite{coherence}.
This measure is basis dependent, as it should be, and has the minimum
value zero, for a diagonal density matrix. However, there is no 
well-defined upper limit to this measure, as it depends on the
dimensionality of the Hilbert space of the system. 
Using this measure, a normalized quantity called coherence was
very recently introduced \cite{nduality}
\begin{equation}
{\mathcal C} = {1\over n-1}\sum_{i\neq j} |\rho_{ij}| ,
\label{coherence}
\end{equation}
where $n$ is the dimensionality of the Hilbert space. This quantity
can assume value between 0 and 1, and can be a measure of wave-nature
just like fringe visibility. Here $\rho_{ij}$ are the matrix elements
of the density operator of the system, in the basis formed by the set
of $n$ orthogonal states, which correspond to the quanton passing through
the $n$ different slits. Based on this new measure of wave nature, the following
duality relation  for n-slit interference was obtained \cite{nduality}
\begin{equation}
{\mathcal D}_Q + {\mathcal C} \le 1.
\label{nduality}
\end{equation}
Here $\mathcal{D}_Q$ is a path distinguishability based on UQSD. The
above is a universal duality relation for n-slit interference. It 
can be shown to reduce to (\ref{2duality}) and (\ref{3duality}) for
two- and three-slit interference, respectively.

However, a shortcoming of the above relation is that it is not clear if
$\mathcal{C}$ can be measured in an experiment. While the fringe visibility
can be experimentally measured to quantify the wave nature in (\ref{2duality}) and
(\ref{3duality}), $\mathcal{C}$ appears to be just a theoretical
construct. The ability to measure quantum coherence is a much sought after
objective, and has been under research attention \cite{coh1,coh2}.
In this paper we show that $\mathcal{C}$  can actually be
measured in a multi-slit interference experiment.

\section{Multi-slit interference and wave nature}
\subsection{Interference visibility}

Let us first look at the case of a $n$-slit quantum interference with 
quantons, without any which-path detection. In $n$-slit interference,
if $|\psi_i\rangle$ is the amplitude
of the quanton to go through the $i$'th slit, then state of
the quanton, after passing through the slit, can be described as sum
of all possible amplitudes, i.e., $|\psi_i\rangle$, to go through different
slits.  Since the slits are well separated, the states $|\psi_i\rangle$
should be orthogonal to each other. We may choose $|\psi_i\rangle$ to be
normalized, and associate a weight factor with it, which determines the
probability of the quanton to go through a particular slit.
\begin{equation}
|\Psi_0\rangle = c_1|\psi_1\rangle + c_2|\psi_2\rangle +
\dots , c_n|\psi_n\rangle .
\label{pure}
\end{equation}
The probability of the particle to go through (say) k'th slit is given by
$|c_k|^2$ and $\sum_{i=1}^n|c_i|^2=1$. Since the states $\{|\psi_i\rangle\}$ are orthonormal, they
can be assumed to form basis states to describe the quanton after it
has passed through the slits.

The interference on the screen is described by the probability density of
particle hitting the screen at particular position
$|\langle x|\Psi_0\rangle|^2$. The expression for the pattern on the
screen will have the following general form
\begin{equation}
|\langle x|\Psi_0\rangle|^2 = \sum_{i=1}^n|c_i|^2|\langle x|\psi_i\rangle|^2 
+ \sum_{j\neq k} c_j^*c_k\langle x|\psi_k\rangle\langle \psi_j|x\rangle.
\label{patternp}
\end{equation}
The first term just represents the sum of patterns formed by the quanton
coming out of individual slits, without any interference. The second term
represents the interference between the amplitudes of quanton coming out
of j'th and k'th slits, summed over all j's and k's.
One would notice here that the multi-slit interference pattern consists of
all possible two-slit terms ($j\neq k$). There are no 3-slit or multiple-slit
terms, which is a direct consquence of the Born rule. This aspect of 
interference has been used to test Born rule in 3-slit interference
experiments \cite{urbasi}.

Now, if one wants to find out which slit the particle went through, one
has to have some kind of path-detecting device in place. Without going into
the details of what kind of device one may use, we just consider certain
fundamental aspects of what such a detection should involve. According to
von Nuemann, in a quantum measurement, the first process should be to let
the detector interact with the quanton and get entangled with it \cite{neumann}.
This involves building up of correlations between the quanton and the
path-detector. The necessary condition for a quantum measurement of 
which slit the quanton went through, is satisfied when each $|\psi_i\rangle$
gets correlated with certain state of the path-detector $|d_i\rangle$.
The combined quanton-detector state, in such a situation, assumes the
following entangled form
\begin{equation}
|\Psi\rangle = c_1|\psi_1\rangle |d_1\rangle + c_2|\psi_2\rangle  |d_2\rangle +
\dots , c_n|\psi_n\rangle |d_n\rangle
\label{ent}
\end{equation}
where $|d_i\rangle$ is the state of the path-detector if the quanton
went through the $i$'th path.
For simplicity, we assume the detector states
$\{|d_i\rangle\}$ to be normalized, but not necessarily orthogonal.
With the path-detector added to the interference setup, the pattern
of the quantons hitting the screen has the following form
\begin{equation}
|\langle x|\Psi_0\rangle|^2 = \sum_{i=1}^n|c_i|^2|\langle x|\psi_i\rangle|^2 
+ \sum_{j\neq k} c_j^*c_k\langle x|\psi_k\rangle\langle \psi_j|x\rangle
\langle d_j|d_k\rangle .
\label{patternr}
\end{equation}
While the first term remains unaffected by the introduction of path-detector,
the second term, which gives rise to interference, is reduced by the
factors $\langle d_j|d_k\rangle$. In fact, for completely orthogonal
path-detector states, $\langle d_j|d_k\rangle = 0$, and the interference
pattern disappears. So, it is clear from this very general analysis that
any attempt to gain information, about which slit the quanton
went through, affects the interference. In other words, probing the 
particle nature of the quanton more precisely, degrades its wave nature
which is characterized by interference.

In order to quantify the wave nature of the quanton, one needs to quantify
the sharpness of the interference pattern. The interference
visibility is defined as
\begin{equation}
\mathcal{V} \equiv \frac{I_{max} - I_{min}}{I_{max} + I_{min}},
\end{equation}
where $I_{max}, I_{min}$ are the maximum and minimum intensity, respectively,
 in some region of the interference pattern. While this visibility
works well for two-slit interference experiments, it is not
clear whether the same definition suffices for multi-slit experiments
\cite{durr,bimonte,englertmb,bimonte1,colesmb}.

\subsection{Coherence as a measure of wave nature}

Coherence as defined by (\ref{coherence}) has been shown to capture 
the wave aspect of a quanton well. Let us calculate the coherence
$\mathcal{C}$ for the state (\ref{ent}).
Since coherence is a
property of the quanton alone, we will first trace out the path-detector
states, to obtain a reduced density matrix of the quanton.
Writing the corresponding density operator for the state (\ref{ent}), and
taking a trace over an orthonormal set of path-detector states, it is 
straightforward to obtain the reduced density operator of the quanton
\begin{equation}
\rho_r = \sum_{k=1}^n\sum_{j=1}^nc_kc_j^* \ \langle d_j|d_k\rangle \ |\psi_k\rangle\langle\psi_j|.
\end{equation}
Inserting this reduced density in the expression for coherence
(\ref{coherence}), and using $\{|\psi_i\rangle\}$ as the basis, one arrives at
\begin{eqnarray}
{\mathcal C} &=& {1\over n-1}\sum_{k\neq j}|\langle\psi_k|\rho_r|\psi_j\rangle|
\nonumber\\
&=& {1\over n-1}\sum_{k\neq j} |c_k||c_j||\langle d_j|d_k\rangle|.
\label{Cn}
\end{eqnarray}
Two extreme cases may be worth noting here.  If the detector states
$\{|d_i\rangle\}$ are all mutually orthogonal to each other, the coherence
$\mathcal{C}$ is 0.  This
implies that in this situation, the fringe visibility goes to
zero. This is also the case when the interference disappears.
On the other extreme, if all the path-detector states $|d_i\rangle$ are
identical, and all the $c_i$'s are equal to $\tfrac{1}{\sqrt{n}}$,
the coherence $\mathcal{C}$ is equal to 1. This is the special case where there is
zero path information about the quanton, and the quanton is equally likely
to pass through any of the $n$ slits. In this situation one gets sharpest
interference. For other nonzero values of $|\langle d_j|d_k\rangle|$,
and different amplitudes of the quanton passing through different slits,
$\mathcal{C}$ will lie between 0 and 1.
\begin{figure}
\centerline{\resizebox{9cm}{!}{\includegraphics{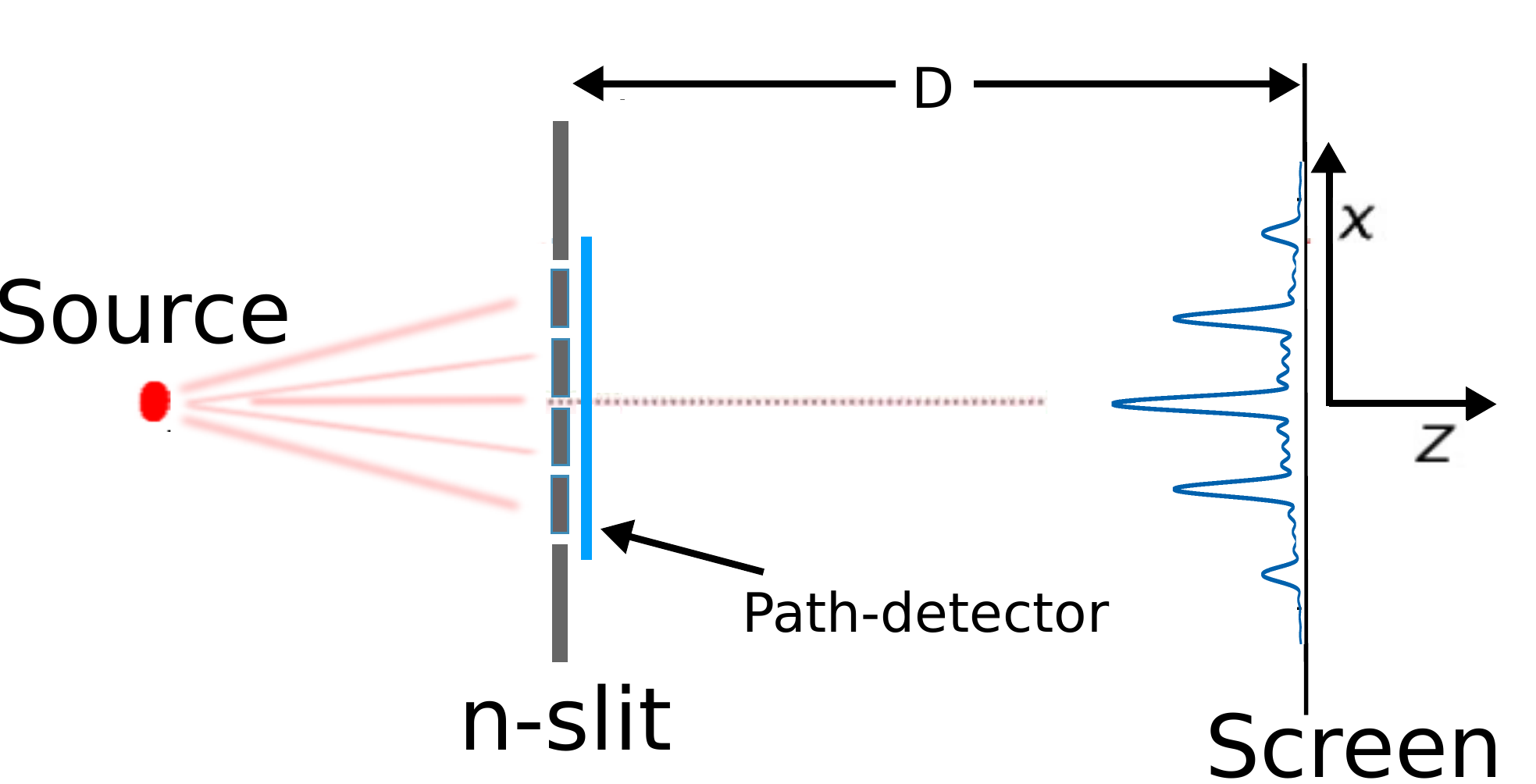}}}
\caption{Schematic diagram of a n-slit interference experiment. 
A quantum path-detector has been added to the setup, which is capable
of obtaining information on which slit the particle passed through.}
\label{nslit}
\end{figure}

D\"{u}rr suggested that any newly defined visibility should satisfy the following
criteria \cite{durr}\\
(1) It should be possible to give a definition of visibility that is
based only on the interference pattern, without explicitly
referring to the matrix elements of $\rho$ .\\
(2) It should vary continuously as a function of the matrix
elements of $\rho$ .\\
(3) If the system shows no interference, visibility should reach its global
minimum.\\
(4) If $\rho$ represents a pure state (i.e., $\rho^2 = \rho$) and all n
beams are equally populated (i.e., all $\rho_{jj}=1/n$), visibility should
reach its global maximum.\\
(5) Visibility considered as a function in the parameter space
($\rho_{11}, \rho_{12},\dots , \rho_{nn}$) should have only global extrema,
no local ones.\\
(6) Visibility should be independent of our choice of the coordinate system.\\
Coherence, as defined by (\ref{Cn}), satisfies D\"{u}rr's criteria (2) through
(6). However, criterion (1) implies that it should be possible to talk
of coherence based only on the interference pattern. This point will be
addressed in the next section.

\section{Measuring Coherence}
\subsection{Pure quanton-detector state}

Next we address the issue of measuring coherence. In order to see how
one may try to measure coherence, one has to have an intereference pattern.
For getting an interference pattern we theoretically analyze the
behavior of a quanton passing through $n$ equally spaced slits. We assume
that the quanton is traveling along the z-axis, and
encounters n-slits which lie in the x-y plane, as shown in Figure \ref{nslit}.
The slits are centered at
$x_j = j\ell$, $j = 1, 2, 3, \dots,~n$. This means that the distance between
any two neighboring slits is $\ell$. We also assume that the state
that emerges from the j'th slit, is a Gaussian along x-axis, centered at
$x = j\ell$. The width of the Gaussian is very small, and is supposed to
be of the order of the width of the slit. Thus the state that emerges from
the slits, at time $t=0$, has the following form, in the position basis
\begin{equation}
\langle x|\Psi(0)\rangle = A\sum_{j=1}^n c_j\exp\left(-\frac{(x-j\ell)^2}{\epsilon^2}\right) |d_j\rangle ,
\label{t0}
\end{equation}
where $A = (2/\pi\epsilon^2)^{1/4}$.
After traveling a distance $D$, in a time $t$, the quanton reaches the screen.
The time evolution of the Gaussian can be calculated by either assuming the
quanton to be particle of mass $m$, moving with a momentum corresponding to a
de Broglie wavelength $\lambda$, or by assuming it to be a photon of 
wavelength $\lambda$ \cite{dillon}. In both cases, the state of the quanton,
at the screen, is given by
\begin{equation}
\langle x|\Psi(t)\rangle = A_t \sum_{j=1}^n c_j\exp\left(-\frac{(x-j\ell)^2}
{\epsilon^2 + i\lambda D/\pi}\right) |d_j\rangle,
\label{t}
\end{equation}
where $A_t = ({2\over\pi(\epsilon+i\lambda D/\pi\epsilon)})^{1/4}$.

The probability density of the quanton, hitting the screen at a position $x$,
can now calculated as
\begin{eqnarray}
|\langle x|\Psi(t)\rangle|^2 &=& |A_t|^2 \sum_{j=1}^n |c_j|^2\exp\left(-\frac{2\epsilon^2(x-j\ell)^2}
{\epsilon^4 + (\lambda D/\pi)^2}\right) \nonumber\\
&&+ \sum_{j\neq k} c_j^*c_k \exp\left(-\frac{(x-j\ell)^2}
{\epsilon^2 - i\lambda D/\pi}\right) \nonumber\\
&&\times \exp\left(-\frac{(x-k\ell)^2}
{\epsilon^2 + i\lambda D/\pi}\right) \langle d_j|d_k\rangle .
\end{eqnarray}
Notice that $\epsilon$, being the width of one slit, is very small, and hence
$\epsilon^4$ is negligible in comparison to $(\lambda D/\pi)^2$. Also, it
is convenient to combine the phases of $c_j$ and $|d_j\rangle$ as
$c_j |d_j\rangle = |c_j||d_j\rangle e^{i\theta_j}$, where $|d_j\rangle$ is
now real.  With these assumptions, the above assumes the following form,
\begin{eqnarray}
|\langle x|\Psi(t)\rangle|^2 &=& |A_t|^2 \sum_{j=1}^n |c_j|^2\exp\left(-\frac{2\epsilon^2(x-j\ell)^2}
{(\lambda D/\pi)^2}\right) \nonumber\\
&&+ \sum_{j\neq k} |c_j||c_k| |\langle d_j|d_k\rangle| \exp\left(-\frac{\epsilon^2 f_{jk}(x)}
{(\lambda D/\pi)^2}\right) \nonumber\\
&&\times~ \cos\left(\frac{2\pi xl(k-j)}{\lambda D} + \frac{\ell^2(j^2-k^2)}
{\lambda D} +(\theta_k-\theta_j)\right) \nonumber\\
\label{intf}
\end{eqnarray}
where $f_{jk}(x) = 2x^2 - 2x\ell(j+k)+(j^2+k^2)\ell^2$. The expression (\ref{intf})
represents a n-slit interference pattern in the presence of path-detectors.
The distance between the primary maxima, or the fringe width is given by
$w = \lambda D/\ell$, which is much larger than the distance between the
two slits, $\ell$, and also the width of a slit $\epsilon$. If the position
on the screen $x$ is on any maximum away from the one at $x=0$, $j\ell$ is
negligible in its comparison. With these things in mind, the probability
density of quantons on the screen can be simplified to
\begin{eqnarray}
|\langle x|\Psi(t)\rangle|^2 &=& |A_t|^2 \sum_{j=1}^n |c_j|^2\exp\left(-\frac{2\epsilon^2x^2}
{(\lambda D/\pi)^2}\right) \nonumber\\
&&+ \sum_{j\neq k} |c_j||c_k| |\langle d_j|d_k\rangle| \exp\left(-\frac{2\epsilon^2 x^2}
{(\lambda D/\pi)^2}\right) \nonumber\\
&&\times~ \cos\left(\frac{2\pi xl(k-j)}{\lambda D} 
 +(\theta_k-\theta_j)\right) .
\label{intfs}
\end{eqnarray}
For simplcity, we assume all the phases to be the same, i.e.,
$\theta_k-\theta_j=0$.  Notice that for $x_m = m\lambda D/\ell,~m=0,1,2,\dots$,
the cosine term is 1, irrespective of the values of $j,k$. These are the
positions of the primary maxima, where the cosine contributions from every 
pair of slits are 1. There are other positions where cosine
terms from some slit pairs are 1, but those from some others are not.
Those are the secondary maxima. The maximum intensity at a primary
maximum is then
given by, $I_{max} = |\langle x_m|\Psi(t)\rangle|^2$, and has the form
\begin{eqnarray}
I_{max} &=& |A_t|^2 \exp\left(-\frac{2\epsilon^2x^2}
{(\lambda D/\pi)^2}\right) \left[ \sum_{j=1}^n|c_j|^2 
+ \sum_{j\neq k} |c_j||c_k| |\langle d_j|d_k\rangle| 
\right] \nonumber\\
\label{imax}
\end{eqnarray}

If the same experiment is performed using incoherent light, instead of coherent light,
one has to average over the phases $\theta_j,\theta_k$ in (\ref{intfs}),
and that would kill the cosine terms. The intensity, at the same position
on the screen as in (\ref{imax}), is then given by
\begin{eqnarray}
I_{inc} &=& |A_t|^2 \exp\left(-\frac{2\epsilon^2x^2}
{(\lambda D/\pi)^2}\right)  \sum_{j=1}^n|c_j|^2  .
\label{iinc}
\end{eqnarray}
One may carry out a careful photon counting experiment with coherent laser
light to measure $I_{max}$. A phase randomizer may then be added to the setup
and the experiment repeated to measure $I_{inc}$. This procedure would allow
one to calculate the following quantity
\begin{eqnarray}
\frac{1}{n-1}\frac{I_{max} - I_{inc}}{I_{inc}} &=& 
\frac{1}{n-1} \sum_{j\neq k} |c_j||c_k| |\langle d_j|d_k\rangle| ,
\label{cmeas}
\end{eqnarray}
where $n$ is the number of slits used in the experiment. Comparing the r.h.s.
of the above equation with (\ref{Cn}), one finds that it is exactly the same
as the coherence $\mathcal{C}$ for the quanton! So, one can measure the
value of coherence in a multislit experiment. Needless to say, since the
experiment involves two parts, it is important to make sure that number
of photons coming out of the slits remains unchanged. Thus the experimentally
measured coherence can be written as
\begin{eqnarray}
\mathcal{C}_{expt} = \frac{1}{n-1}\frac{I_{max} - I_{inc}}{I_{inc}} .
\label{cexpt}
\end{eqnarray}
In real situations, the interference fringes occur within a Gaussian envelope,
and the maxima on either side of the Gaussian peak will be gradually lower
in intensity. However, it does not matter which primary maximum is chosen
for measuring the intensity. The procedure makes sure that the Gaussian part cancels
out and one gets an expression for coherence (\ref{cmeas}) independent of
the position of the primary maximum. With the measured coherence given by
(\ref{cexpt}), $\mathcal{C}$ also satisfies D\"{u}rr's criterion (1) for
being a good measure of interference visibility.

\subsection{Mixed quanton-detector state}

In real life situations, it may happen that the quanton is affected by
the environment, and can no longer be described by a pure state. In such
situations, the state of the quanton and path-detector combined, may be
represented by a mixed state density operator
\begin{equation}
\rho_m = \sum_{j=1}^n\sum_{k=1}^n q_{jk} |\psi_j\rangle\langle\psi_k|\otimes
|d_j\rangle\langle d_k| ,
\end{equation}
where $q_{jk}$ are complex numbers. In this case, coherence is given by 
\cite{nduality}
\begin{eqnarray}
{\mathcal C} &=& 
{1\over n-1}\sum_{k\neq j} |q_{jk}||\langle d_k|d_j\rangle|.
\label{Cnm}
\end{eqnarray}
One can then follow a procedure closely similar to that in the preceding 
subsection, and obtain the reduced density matrix for the quanton at the screen.
The diagonal part of the density matrix in the position representation, is
then given by
\begin{eqnarray}
\rho_{mr}(x,x,t) &=& |A_t|^2 \sum_{j=1}^n |q_{jj}|\exp\left(-\frac{2\epsilon^2x^2}
{(\lambda D/\pi)^2}\right) \nonumber\\
&&+ \sum_{j\neq k} |q_{jk}| |\langle d_j|d_k\rangle| \exp\left(-\frac{2\epsilon^2 x^2}
{(\lambda D/\pi)^2}\right) \nonumber\\
&&\times~ \cos\left(\frac{2\pi xl(k-j)}{\lambda D} 
 +\phi_{jk}\right) .
\label{intfm}
\end{eqnarray}
where $q_{jk}\langle d_j|d_k\rangle=|q_{jk}| |\langle d_j|d_k\rangle|e^{i\phi_{jk}}$.
The maximum intensity at a primary maximum is then given by
$I_{max} = \rho_{mr}(x_m,x_m,t)$. Again, for incoherent light, random variations
in $\phi_{jk}$ will kill the cosine terms. Experimentally measured coherence is
then given by
\begin{eqnarray}
\mathcal{C}_{expt} = \frac{1}{n-1}\frac{I_{max} - I_{inc}}{I_{inc}} &=& 
\frac{1}{n-1} \sum_{j\neq k} |q_{jk}| |\langle d_j|d_k\rangle|, \nonumber\\
\label{cmeasm}
\end{eqnarray}
which agrees with the theoretical expression for coherence (\ref{Cnm}).

\subsection{Coherence of the incoming quanton}

In the preceding subsections, we looked at the problem of measuring the
coherence
of a quanton as it emerges from n slits, and a path detector tries to get
information about which of the n slits the quanton passed through. The
coherence of the quanton degrades in the process of path detection.
This procedure is well suited for testing wave-particle duality relations.
On the other hand, if one is interested in measuring the coherence of the 
incoming quanton as it enters the slits, the above procedure may not be well
suited for the job. One reason is that in the preceding procedure,
one needs to change the incoming state by randomizing the phases at
different slits. One may want a measuring procedure where the incoming state
is not disturbed. Another reason could be that the phase randomizer may
not be easily realizable.
We propose another procedure for this particular case.

We assume that we have a n-slit system and a path detector in place where
the path-distinguishability is tunable. At the least, it should be switchable 
between two modes corresponding to (a) making all the paths completely
{\em indistinguishable} and (b) making all the paths fully
{\em distinguishable}. We denote the the two cases (a) and (b) by 
$\parallel$ and $\perp$, respectively. First the intensity at a
primary maximum is measured when all the n paths are indistinguishable,
i.e., $|d_i\rangle$s are all identical and parallel. For a pure quanton
state, this intensity can be obtained simply by putting
$|\langle d_j|d_k\rangle|=1$ for all $j,k$ in (\ref{imax}), and  is given by
\begin{eqnarray}
I_{max}^{\parallel} &=& |A_t|^2 \exp\left(-\frac{2\epsilon^2x^2}
{(\lambda D/\pi)^2}\right) \left[ \sum_{j=1}^n|c_j|^2 
+ \sum_{j\neq k} |c_j||c_k| 
\right] .
\label{iparall}
\end{eqnarray}
Next the path-detector is switched to the mode where all the n paths
are fully distinguishable, and the intensity is measured at the same
position on the screen as before. This intentsity can be obtained simply
by putting $\langle d_j|d_k\rangle=0$ for all $j\neq k$ in (\ref{imax}),
and  is given by
\begin{eqnarray}
I_{max}^{\perp} &=& |A_t|^2 \exp\left(-\frac{2\epsilon^2x^2}
{(\lambda D/\pi)^2}\right) \sum_{j=1}^n|c_j|^2 .
\label{iperp}
\end{eqnarray}
Coherence of the incoming quanton can then be measured as
\begin{eqnarray}
\mathcal{C}_{expt}^{0} = \frac{1}{n-1}\frac{I_{max}^{\parallel} - I_{max}^{\perp}}{I_{max}^{\perp}} &=& 
\frac{1}{n-1} \sum_{j\neq k} |c_j||c_k|  . 
\label{cinput}
\end{eqnarray}
The right hand side of the above is the coherence of the incoming quanton,
in the basis of the n-slit paths, {\em unmodified by the path-detector}.
Doing away with the factor $\frac{1}{n-1}$ in the above, gives the
unnormalized coherence introduced by Baumgratz et.al. \cite{coherence}.
Exactly the same procedure will also work for quantons described by a
mixed state.

So, in order to measure the coherence of a beam of light or massive particles,
one needs to introduce a n-slit setup in its path, and measure the intensity
at a primary maximum, in two different modes of the path-detector.

\subsection{Path distinguishability}

For completeness, we briefly discuss the path distinguishability 
$\mathcal{D}_Q$ introduced in \cite{nduality}.
From (\ref{ent}) it is clear that in order to tell which slit the 
quanton went through, one has to be able to tell which path-detector
state, out of the set $\{|d_j\rangle\}$ materialized.
If one is only interested in the path-dector and is not concerned with what
happens to the quanton after passing through the slits, one can write
the density operator corresponding to the state (\ref{ent}), and trace
out the states of the quanton. The resulting reduced density operator
for the path detector is
\begin{equation}
\rho_d = \sum_{j=1}^n |c_j|^2 |d_j\rangle\langle d_j|.
\end{equation}
So, essentially a particular detector state $|d_k\rangle$ occurs with a
probability $|c_k|^2$. The only problem is that the states $\{|d_j\rangle\}$
are not orthogonal to each other, in general. The problem of path-detection
then boils down to the following. Certain known non-orthogonal states
$|d_k\rangle$ occur at random, one at a time, and which one occurs is
not known. One has to tell {\em unambiguously} which of the set
$\{|d_j\rangle\}$ is a given state. Our view is that the answer has to be
unambiguous, only then can one claim that the particle has gone through
a particular slit. An answer which is ambiguous, doesn't serve the purpose.

In general, the best strategy to distinguish between non-orthogonal states
is unambiguous quantum state discrimination (UQSD)
\cite{uqsd,dieks,peres,jaeger2,bergou}. In UQSD, the maximum probability
with which non-orthogonal pure states can be {\em unambiguously} distinguished,
can be calculated \cite{zhang,qiu}. The probability of successfully distinguishing
between $n$ non-othogonal states $\{|d_j\rangle\}$, via UQSD is bounded by \cite{zhang,qiu}
\begin{equation}
P_n \le 1 - {1\over n-1}\sum_{k\neq j} |c_k||c_j| |\langle d_k|d_j\rangle|.
\label{pn}
\end{equation}
It should be reiterated that the probabilities $|c_j|^2$ are decided by the initial superposition in the quanton state.
The path-distinguishability ${\mathcal D}_Q$ can then be {\em defined} as
the maximum
probability with which the $n$ paths of the particle can be distinguished
without any error, and is given here by the upper bound of (\ref{pn}) \cite{nduality} 
\begin{equation}
{\mathcal D}_Q \equiv 1 - {1\over n-1}\sum_{i\neq j} |c_i||c_j| |\langle d_i|d_j\rangle|.
\label{D}
\end{equation}
The path-distinguishability can take values between 0 and 1. For all mutually orthogonal $\{|d_j\rangle\}$, one gets ${\mathcal D}_Q=1$. It should be
mentioned that that UQSD works for linearly independent states. In the
situation where the set $\{|d_j\rangle\}$ is not linearly independent, UQSD
will not work, and one has to employ some other strategy to unambiguously
tell, probably in a limited way, which of $\{|d_j\rangle\}$ states one has
obtained. However, one can still continue to use the upper bound given by
(\ref{pn}). Of course in this scenario, the bound may not be reachable.

The quantum coherence $\mathcal{C}$, given by (\ref{Cn}) or (\ref{Cnm}),
and the path distinguishability, given by (\ref{D}), are bounded by the
inequality (\ref{nduality}) \cite{nduality}. In the light of the 
demonstration of experimental measurability of $\mathcal{C}$, the
inequality (\ref{nduality}) can be treated as a general duality relation
which works for interference experiments with any number of slits.

\section{Conclusions \label{sec:concl}}

In conclusion, we have presented a method of experimentally measuring 
the recently introduced measure of coherence, $\mathcal{C}$ in a
multi-slit interference experiment. Coherence $\mathcal{C}$  had earlier
been argued to be a good candidate for quantifying the wave-nature of
quantons in interference experiments, and a duality relation between ${\mathcal{C}}$ 
and the path-distinguishability $\mathcal{D}_Q$, based on UQSD, had been
proved \cite{nduality}. Here $\mathcal{C}$ has been shown to be related to measured
intensities in interference experiments, which puts it firmly as a
good measure of wave nature of quantons in multi-slit interference experiments.
Additionally, it makes $\mathcal{D}_Q + \mathcal{C} \le 1$ a universal
wave-particle duality relation, valid for interference experiments with
any number of slits.

Apart from the issue of wave-particle duality, a method of measuring the
coherence of any given beam of quantons has also been proposed.

\section*{Acknowledgement}
T.P. is thankful to the Centre for Theoretical Physics at Jamia Millia
Islamia for providing the facilities of the Centre during the course of 
this work. Authors thank an anonymous referee for suggesting an extension
of the results.

\end{document}